# Unsuspected implications arising from assumptions in simulations: Insights from recasting a forest growth model in system dynamics

Jerome K Vanclay

Forest Research Centre, Southern Cross University

PO Box 157, Lismore NSW 2480, Australia

*Abstract*

Familiarity with a simulation platform can seduce modellers into accepting untested assumptions for convenience of implementation. These assumptions may have consequences greater than commonly suspected, and it is important that modellers remain mindful of assumptions and remain diligent with sensitivity testing.

Familiarity with a technique can lead to complacency, and alternative approaches and software can reveal untested assumptions. Visual modelling environments based on system dynamics may help to make critical assumptions more evident by offering an accessible visual overview and empowering a focus on representational rather than computational efficiency. This capacity is illustrated using a cohort-based forest growth model developed for mixed species forest.

The alternative model implementation revealed that untested assumptions in the original model could have substantial influence on simulated outcomes.

An important implication is that modellers should remain conscious of all assumptions, consider alternative implementations that reveal assumptions more clearly, and conduct sensitivity tests to inform decisions.

*1. Introduction*

Modellers generally try to implement ecological concepts faithfully and completely, but there is inevitably a tendency for model implementation to be influenced by technology, both hardware and software. This tendency is often subtle, with small artefacts introduced into a simulator with little discussion, deemed prudent to achieve efficiency or expediency. Whilst modelling, like management, is usually a compromise to deliver timely and useful results, the danger of technological limitations is that they are often subtle, understated and untested. A search of the literature suggests that authors are more likely to discuss compromises made because of data limitations than those introduced because of software limitations, perhaps because of the ease of applying a familiar approach (i.e., a 'golden hammer', Brown *et al.,* 1998). This is somewhat worrisome, because minor compromises introduced for convenience in implementation may introduce substantial and unsuspected effects into simulated predictions. Where such consequences are suspected, they may be remedied during model evaluation, but two dangers remain: firstly that many modellers do not suspect the extent to which their chosen software influences their implementation or the magnitude of the consequences that may arise; and as a result they tend not to discuss these assumptions in simulator documentation. Discussions informing this paper suggest that many modellers discount and dismiss these dangers and their consequences too readily, despite empirical evidence of their importance. This paper seeks to alert modellers to these dangers and to encourage greater reporting of assumptions and compromises in both model design and in their implementation in simulators (*sensu* Pretzsch et al 2002).

This paper explores some aspects of compromises made during the implementation of a simulator, drawing on a published growth model for north Queensland rainforests (Vanclay, 1989a). The most-widely used version of this simulator was written in standard Fortran-77 and executed under Unix on a Vax 11-780 computer system, a platform relatively free of limitations and well understood by many modellers. This simulator was not naïve implementation, but was a deliberate decision informed by familiarity with Wirth's (1985) work on the importance of data structures, by knowledge of diverse programming languages (e.g., Fortran, Pascal, Simula, and Simscript), by feedback received through earlier publication of several forest growth models, and by familiarity with other modelling approaches (Vanclay, 1983). The model was widely emulated (e.g., Ong and Kleine, 1996; Alder and Silva, 2000), and was instrumental in informing changes in forest management (Preston and Vanclay, 1988; Vanclay, 1996a). But a re-evaluation of the simulator as part of an undergraduate teaching program has revealed that some untested assumptions implicit in the Fortran implementation become more evident when the model is implemented in other platforms. This paper draws attention to two issues important in modelling: the need for critical review of assumptions implicit in model implementation, and the utility of considering alternative formulations to encourage explicit recognition of such assumptions.

*2. Literature*

Early models in the form of alignment charts (e.g., Reineke, 1927) were transparent in their application, if not their development, but the advent of statistical and computer-based models (e.g., Buckman, 1962; Clutter, 1963) commenced a subtle trend of incomplete disclosure of simulator details in published documentation. Whilst this is generally not deliberate, text-based publication of component functional relationships is rarely sufficient to reproduce a simulator (Robinson and

Monserud, 2003), because many details intimate to a particular implementation may affect predictions (Villa et al., 2009). Guidelines (e.g., Pretzsch *et al.,* 2002) have improved the standard of documentation, but there are limits on the extent to which text-based descriptions can adequately and efficiently describe computer-based implementations. Visual 'icon-based' modelling environments, typically based on system dynamics, can help to reveal model structure and reduce the 'black box' syndrome (Smith *et al*., 2005). One key advantage of these visual modelling environments is that the diagram is both the model and the simulator, unlike some early applications where system dynamics flowchart was merely documentation about the model (e.g., Kalgraf 1979). Unfortunately, the potential of these new tools for model design and simulator function has received little attention. Researchers have investigated how simulation results may depend on the functional form of tree growth equations (e.g., Elkin *et al.*, 2012), but have not adequately addressed how simulators may also depend on host software used to implement the model. It is surprising that these potential dependencies have not been researched, because several popular modelling platforms have different unique features. For instance, Stella (Doerr, 1996; Eskrootchi and Oskrochi, 2010) has a conveyor-stock that handles time-lags efficiently (e.g., Blackwell *et al*., 2001), whilst Simile (Muetzelfeldt and Massheder, 2003; Muetzelfeldt, 2010) offers multiple-instance submodels that facilitate individual-based modelling, so it seems likely that these features may influence the implementation of models on respective platforms.

Vanclay's (1989a) model for simulating growth and yield of tropical rainforest is widely cited and has provided the inspiration for several other forest growth simulators still in use today (e.g., Ong and Kleine, 1996; Alder and Silva, 2000), but the traditional journal article presentation of this model remains silent about some key details concerning the design and implementation of this simulator. Although this model has been superseded (Vanclay, 1994a), it retains considerable utility for teaching because of its relative simplicity. An alternative implementation of this simulator using the visual modelling environment Simile is revealing, informative and more pedagogic than the original Fortran code and journal article. Although visual system-dynamics modelling tools have been available and used increasingly for over two decades (Bossel, 1991; Doerr, 1996; Garcia, 2013), their role in informing and sharing information remains underutilized. This paper discusses new insights offered through the Simile presentation of this model, and refutes the assertion (Dufour-Kowalski *et al.,* 2012) that visual modelling environments such as Simile are not well suited for forest growth modelling.

The best way to share information about a model is to share the simulator itself, in a form that is open-source and implemented in a generic easily-understood way. However, many useful computer languages are accessible only to a small number of practitioners, and important details of simulators coded in these languages may be hidden like the proverbial 'needle in a haystack' (Muetzelfeldt, 2004; Villa *et al.,* 2009). Thus there is limited benefit in sharing proprietary computer code, unless the simulator is well documented and the language widely utilized and freely available. These hidden aspects of computer-based simulators may well be a 'skeleton in the closet': many experienced modellers have anecdotes about the importance of untested 'calibration factors' or about unintended consequences of particular data structures (e.g., Hamilton, 1994), but these are rarely documented in the formal literature, allowing these oversights to be perpetuated.

Muetzelfeldt's interest in dynamic representation of ecological relationships (e.g., Muetzelfeldt *et al.*, 1989) led to the development of Simile, a visual modelling environment useful for modelling ecological and agricultural systems (Muetzelfeldt and Taylor, 1997; Muetzelfeldt and Massheder, 2003). Simile employs a declarative modelling approach and saves models as structured text files, able to be processed by other software (Muetzelfeldt, 2004). It offers several constructs useful for modelling plant growth and related concepts (Prabhu *et al.*, 2003; Vanclay, 2006a), as well as more abstract issues (e.g., Haggith *et al.*, 2003). Thus Simile provides an interesting vehicle to illuminate, compare and share existing models (Davey *et al.*, 2009). Simile is not unique in this ability, and Vensim is an alternative that has been used to implement forest growth simulators (e.g., Garcia and Ruiz, 2003).

Rainforests provide an informative case study because many of their characteristics pose challenges in modelling: the large number of species, the wide range of stem sizes and growth patterns, and the paucity of calibration data all amplify inherent challenges (Vanclay, 1991d; Clark and Clark, 1999; Picard and Franc, 2003). Vanclay's (1989a) model for Queensland rainforests is a frequently cited example, and illustrates several insights that may be gained by recasting a simulator from its original Fortran into a visual system dynamics representation.

*3. Recasting the model in Simile*

The essence of Vanclay's (1989a) model is summarised in the Simile model illustrated in Figure 1. Input data comprise a series of triplets reflecting species, size (diameter at breast height, 1.3 m above ground) and stocking (N/ha) of each tree measured on a field plot. The symbols used in Figure 1 are standard system dynamics notation common to several visual modelling environments (Ford 1999), and are not merely a diagram, but form a model with which simulations may be made. Rectangles (▬) are stocks or *compartments* that represent an amount of substance (numbers of trees, size of a tree), and change only via a *flow* (⤴, e.g., die, grow). Clouds (☁) denote exogenous elements of no further use in the model. Variables (⬤) may contain data, equations or lookup tables. Submodels can represent a single instance (▭) usually to provide context to other elements, or may represent multiple instances (▭) such as a list of trees (cf. an array).

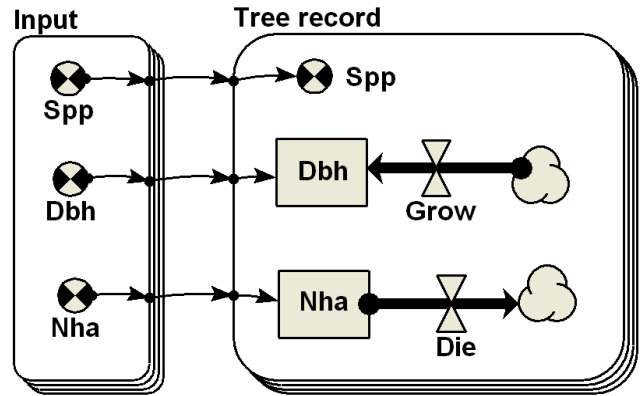

**Figure 1**. Simile diagram of the underlying philosophy of Vanclay's (1989a) model.

The elements of Figure 1 represent the key information needed for individual-based modelling of forests (Weiskittel *et al.*, 2011). These data may be read from file by the growth model, with each input record forming one of many tree records or cohorts in the model, a technique used widely in forest growth modelling (Vanclay, 1994b; Porte and Bartelink, 2002). These cohorts retain the species identity, and progressively increase tree size to reflect growth, and reduce stocking to reflect mortality.

Figure 1 is not merely a diagram, but when created in Simile, is the simulator that runs and creates simulated outputs. One of the strengths of Simile is that the user interface simultaneously creates a diagrammatic overview, the executable code, and model documentation (a 'mouseover' displays any comment included with each symbol).

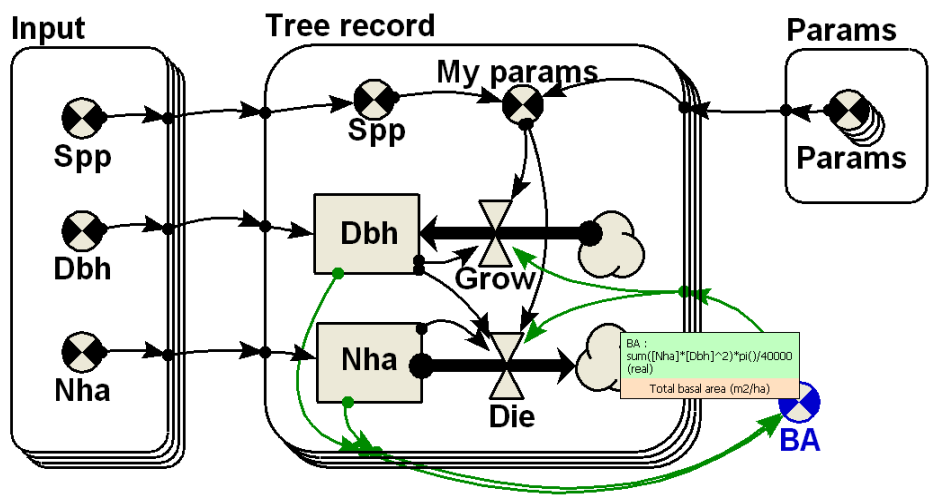

**Figure 2**. Growth and mortality depend on tree size and stand density. Equation parameters are drawn from an external file. Mouseover causes icon contents (the formula for calculating stand basal area) to be displayed.

Figure 2 contains additional detail to show how coefficients for equations are drawn from file, specific to each of a small number of species groups (e.g., Köhler and Huth, 1998). Stand basal area (*BA*) is computed from the size and stocking represented by each tree record. A key concept is that *BA* is external to the *Tree record* submodel and can 'see' all the diameters, and thus computes the stand-level statistic, whereas *Grow* sees only the single *Dbh* within its own *Tree record*. Figure 2 also adds influences that denote how both growth and mortality depend on tree size and stand density. Thus although Figure 2 is a crude and partial approximation of Vanclay's (1989a) model, it gives a useful overview of much model detail and introduces concepts of Simile notation, most of which is standard system-dynamics notation.

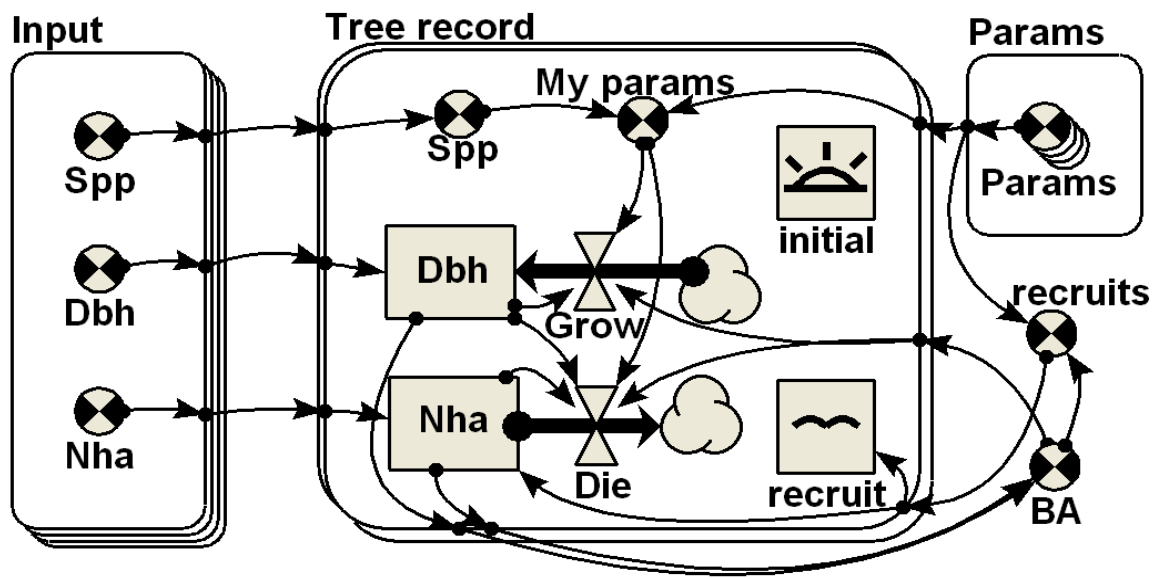

**Figure 3**. Recruitment is density-dependent, and initiates a new *Tree record*.

Figure 3 adds recruitment to the model using the migration symbol (⁀). Because recruitment adds additional tree records, the *Tree record* submodel is no longer a list with fixed membership (like the input file with multiple outlines bottom right), but is denoted as a population (denoted by multiple outlines top left) in which the number of records can increase or decrease during the simulation. Thus the sunrise symbol (*initial*) indicates the initial membership of the population, and the migration

(⁀) symbol (*recruit*) initiates the new records needed to simulate recruitment. The number of *recruits* is estimated from stand density. Whilst Figure 3 is a slightly simplified representation of Vanclay's (1989a) model, it serves to explain several details of Simile in an accessible way. Readers seeking more specifics about other simulation constructs available in Simile are directed to Simulistics (2013).

A reviewer asked whether the migration symbol (🕊) could be changed from a bird to an icon that better represents tree dynamics: it can be, but such customization does not facilitate communication. To take an example from Microsoft: a user may not like the scissors icon used to denote 'cut' in Microsoft software, but everyone understands what this icon denotes, and shared understanding should take precedence over personal aesthetics.

One aspect of many forest growth simulators that is often poorly documented is that of record splitting, a technique introduced to emulate growth variation, and avoid bias associated with Jensen's inequality (Jensen, 1906; Duursma and Robinson, 2003). This is an established technique often known as a 'swindle' (Jacoby and Harrison, 1962; Stage, 1973; Simon, 1976; Hedayat and Su, 2012) employed in many simulators either as binary splitting (e.g., Alder and Silva, 2000; Ledermann and Eckmüllner, 2004) or as tripling (e.g., Stage, 1973; Crookston and Dixon, 2005) to attain more realistic projections, by simulating heterogeneous growth deterministically without the overheads of stochastic simulation. Despite the care with which modellers craft this splitting into their source code, the resulting effects are rarely tested exhaustively, and may overlook serial correlation of growth (often handled implicitly via an estimate of one-sided competition, e.g., Cole and Stage, 1972). This inequality has the potential to introduce considerable bias into estimates (Valle, 2011). Simile provides an icon ( ⬤ ) to facilitate record splitting and inheritance of attributes (Figure 4) that facilitates exploration of this issue.

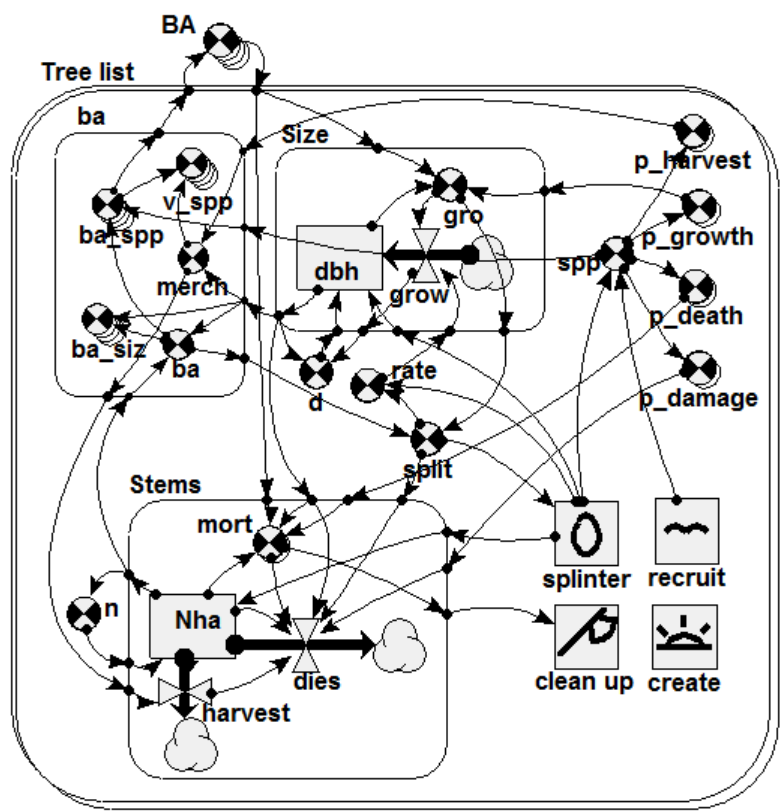

**Figure 4**. The *Tree list* submodel central to the Simile representation of Vanclay's (1989a) model.

Figure 4 shows the major part of the Simile implementation of Vanclay's (1989a) model. The *ba* submodel at top left simply tabulates the basal area and volume per tree by size class and by species; it provides a collation role and serves no functional purpose. The series of variables at top right extract the relevant species-specific parameters for use in other functions. Four symbols at bottom right influence the number of records in the tree list, creating the initial number (⌣), adding recruitment (⌒), splitting records ( ● ), and removing redundant records (🔑) when they represent an infinitesimal stocking. The key functional components of this submodel lie within the *Size* and *Stems* submodels which estimate diameter increment and mortality respectively. Between these are two variables, *rate* and *split*, that deal with serial correlation of increments and record splitting respectively. When a large cohort is split into two smaller cohorts with faster and slower than average growth, *rate* maintains this growth difference to preserve the serial correlation specified by the user.

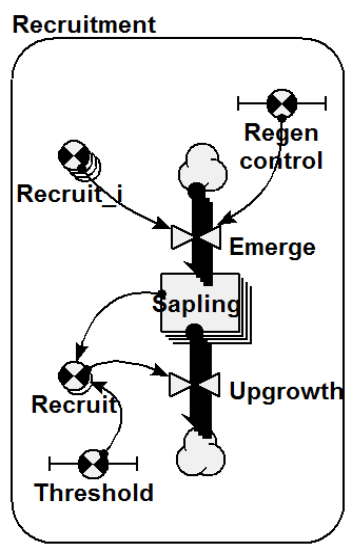

**Figure 5**. Recruitment submodel simulates regeneration and recruitment.

Figure 5 illustrates the recruitment submodel, which predicts regeneration within each species group, standardises it to the total expected recruitment (Vanclay, 1989a), and adds the stem count to a tally of saplings (by species class). When one of these counts exceeds a user-specified threshold, recruitment to the main model is initiated. This *Recruitment* submodel is functionally similar, but does not exactly replicate the algorithm in the original Fortran model, which initiated five new records at each time step, and subsequently amalgamated them during the process of rationalizing the *Tree list*. The approach illustrated in Figure 5 accumulates saplings until one of the species groups exceeds the user-specified threshold of abundance, and then transfers members of that species group to the main tree list. This approach minimizes the housekeeping required within the main *Tree list* submodel, and reproduces the commonly-observed tendency for rainforest recruitment to be clustered in time and space (e.g., Okuda *et al.,* 1997; Connell and Green, 2000; Nebel *et al.,* 2001; Kariuki *et al.,* 2006).

This alterative way to process recruitment illustrates one of the benefits of recasting models: advances in computer technology have created more revealing ways to represent concepts, and pose fewer restrictions on computational resources, allowing modellers to focus on good representations rather than computational efficiency (Vanclay, 2003). The challenge is for modellers to break free of old paradigms and fully exploit these new opportunities, and not merely follow a well-trodden path without question. Leary (1985, p.47), commenting on a related matter, warned that "what began as an interim solution (site index) to a difficult problem (geocentric approach [to site productivity estimation]) should not now be called the solution to the original problem" and his caveat applies equally to modelling.

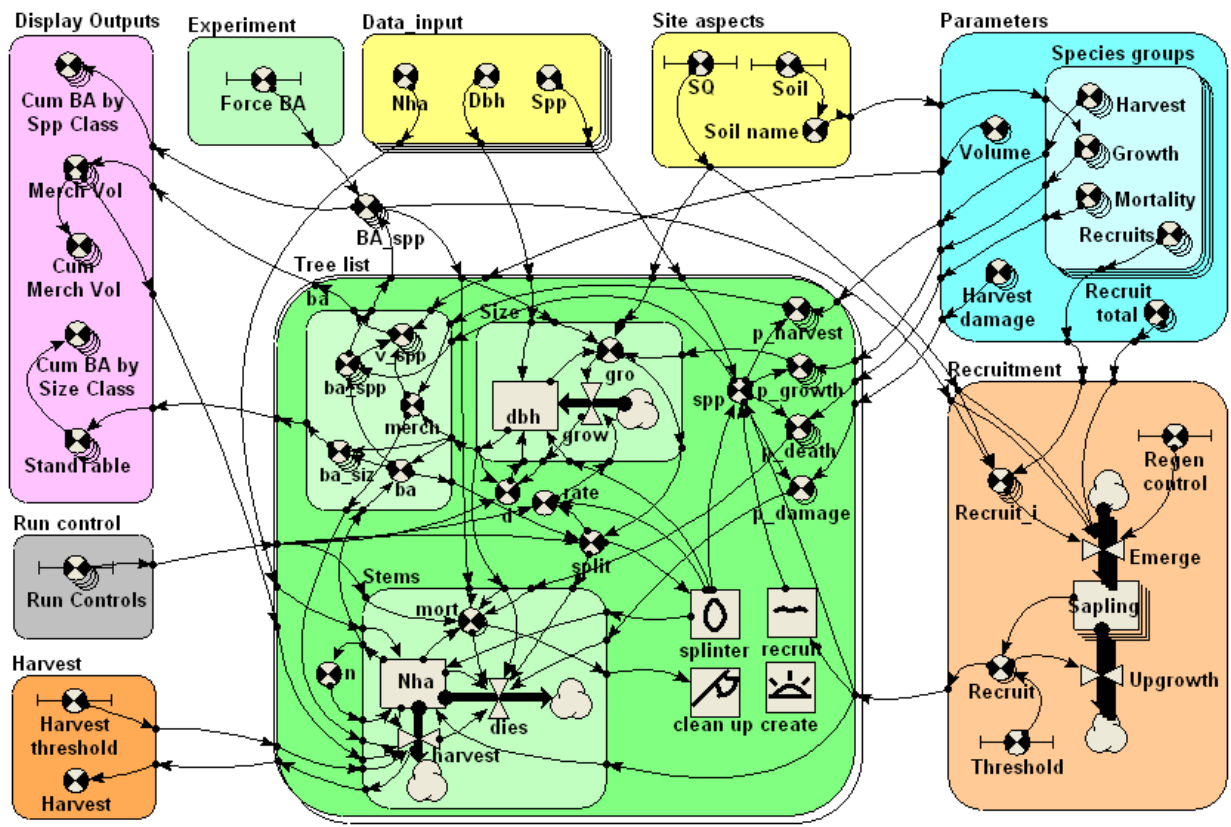

**Figure 6**. The complete model of Vanclay (1989a) implemented in Simile, with enhancements.

Figure 6 illustrates the complete model, a Simile equivalent of Vanclay's (1989a) Fortran simulator, which differs from the original in only a few minor respects. The *Run control* submodel (left, middle) contains a slider that enables users to adjust parameters that were deeply embedded in the original Fortran code, that control the frequency and variance of record splitting, the serial correlation of increments, and the granularity of the simulator. The ability to vary granularity was not available in Vanclay's (1989a) model which was aimed at estate-level planning, and was implemented only later (Vanclay, 1991a). Granularity deals with the 'lumpiness' (cf. sugar lumps versus table sugar) of outputs and the use of stochastic functions. If a user stipulates a coarse granularity (e.g., 1.0), the simulator behaves as a single tree model with stochastic mortality of stems with frequency <1/ha so that individual tree records always represent at least 1 tree per hectare. In contrast, fine granularity (e.g., <0.01) ensures deterministic outputs with a resolution characteristic of estate-level averages.

In the original Fortran simulator, record splitting was hard-coded to assign 1.3 times the average increment to a quarter of the stems, whilst the remaining stems were assigned 0.9 times the predicted increment. In Figure 6, these growth rates are under user control as an easily-accessible slider. The conditions for initiating record splitting and the degree of serial correlation of increments were hidden deep in the original Fortran code, but are under user control as a slider in Figure 6. These controls could be shown explicitly as separate sliders, but for compactness are here compressed into a single vector.

The *Parameters* submodel (Figure 6, top right) contains parameters for diameter increment (Anon, 1987), for mortality and recruitment (Vanclay, 1989a), for volume estimation (Anon, 1981), and for harvesting and logging damage (Preston and Vanclay, 1988). The practice of collating all parameters into one submodel (rather than scattering them throughout the model) facilitates verification and maintenance of the simulator. The species grouping used in this model differs slightly from Vanclay (1989a) in using the same species groups for growth, harvesting and volume estimation, whereas Vanclay's (1989a) model used a 3-digit species code, with each digit denoting the volume, harvesting or growth equation to be applied. These differences in the species grouping are minor, and the simplification employed here makes little practical difference and improves clarity for teaching.

The Simile implementation of this model encourages exploration of simulator assumptions not possible in the original version. For instance, a slider labelled *Force BA* (Figure 6 top left) allows users to over-ride the calculated stand basal area and thus to explore growth patterns and possible thinning regimes by holding standing basal area constant. Such experimentation is easy to implement within Simile, and can offer new insights and better understanding of a simulator.

*4. Results*

Alternative strategies of record splitting may have considerable influence on predictions. Whilst differences may appear rather small over a decade, they can accumulate to contribute substantial differences that may dramatically affect the timing of harvests which is typically influenced by the number of stems exceeding some threshold. This difference does not eventuate in the case of a single stem simulated as an individual tree model (granularity = 1), but can be substantial when cohorts of post-disturbance regeneration are simulated at the estate level (granularity <0.01).

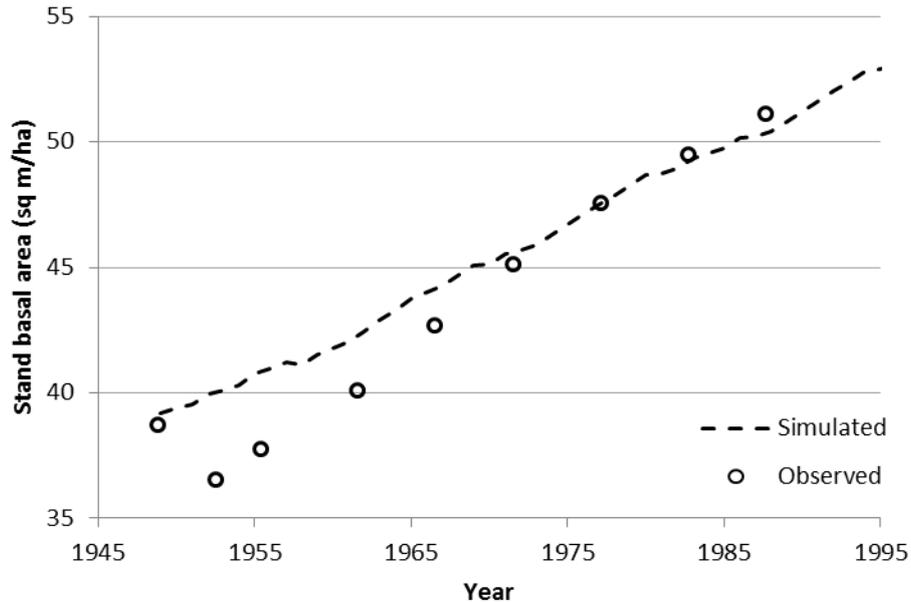

**Figure 7**. Predicted growth of forest in Experiment 78 Plot 1, showing observed and predicted basal areas during a 50-year period from the initial measurement in 1948.

It is no longer possible to demonstrate a side-by-side comparison of the Fortran and Simile implementations of this model, because of agency restrictions and subsequent loss of the original source code (Vanclay 2006b). Whilst it would be possible to reconstruct a Fortran implementation from extant documentation, such reconstruction is unlikely to recreate an unbiased implementation given prior knowledge that the reconstruction would be used test for inadequate assumptions. However, it is possible to demonstate that the Simile implementation offers reasonable growth predictions: Figure 7 illustrates that a 50-year simulation compares favourably with growth observations on one of the plots with the longest measurement history (Experiment 78, Plot 1, last harvested in 1943). This plot was selected for this and subsequent comparisons because of the long duration of observation, the absence of disturbance, and the frequency of remeasures. Whilst Figure 7 is insufficient to serve as a formal evaluation, it serves as a simple reality check prior to the current question surrounding the influence of implicit assumptions in constructing the simulator.

The simulator configuration presented in Figure 6 simplifies the testing of several assumptions: the threshold number of saplings for recruitment to be initiated, the magnitude of the swindle (i.e., the variance of growth rates), the duration of serial correlation in growth rates, the threshold difference for a cohort to be split, and the granularity of mortality (i.e., whether it emulates the individual tree or estate-level average). Most of these exhibit little sensitivity, and substantial changes in these parameters barely influence predictions likely to influence forest management – but two of these parameters have a substantial effect in the longer term: the magnitude of the swindle and the duration of serial correlation.

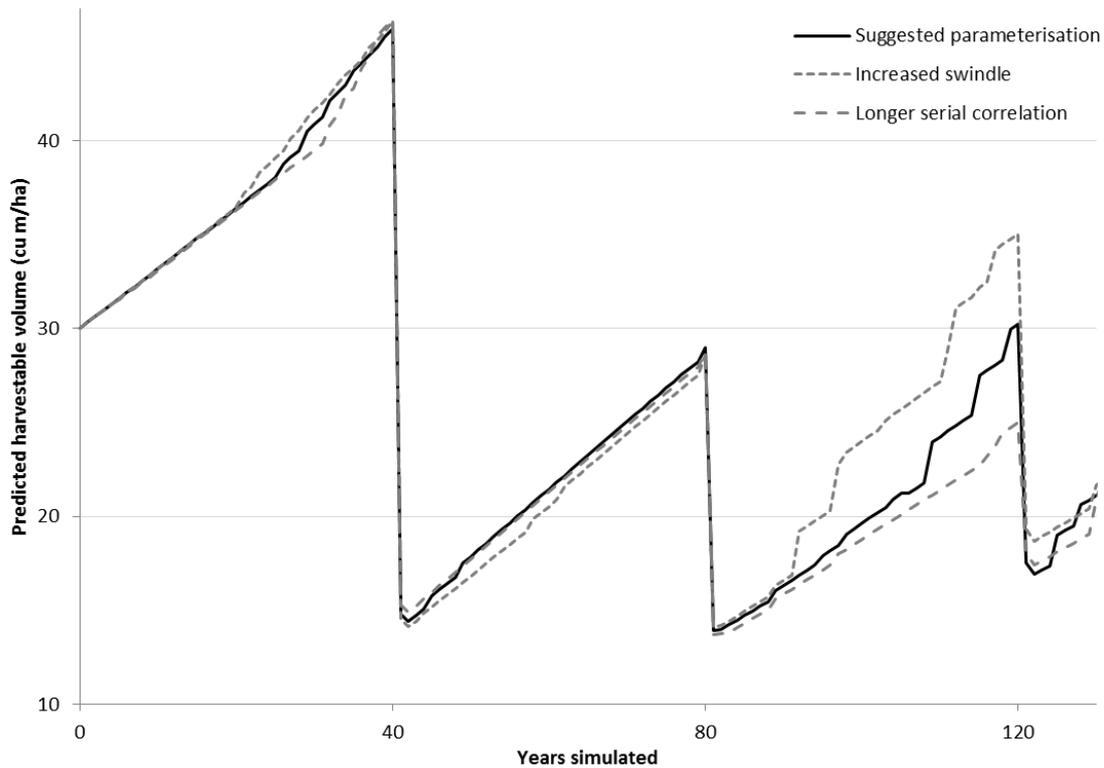

**Figure 8**. Predicted timber harvest during three 40-year harvesting cycles under three different assumptions. Note the difference in the predicted harvest during the third cycle.

Figure 8 illustrates that modest changes in these parameters cause substantial changes (i.e. ±20%) in the predicted volume available during the 3$^{rd}$ successive harvest with a nominal 40-year interval. This 120-year duration was adopted as the basis of comparison, because at the time the model was developed during the 1980s, three successive non-declining harvests were commonly accepted as a criterion for sustainability (Higgins, 1977; Preston and Vanclay, 1988; Botkin, 1993; Vanclay, 1996b). Ironically, the harvest predicted for the third cycle depends heavily on untested assumptions regarding the swindle and serial correlation (Figure 8), both of which are rarely examined during the commissioning of a simulator. The consequences are counter-intuitive: one might expect that serial correlation might lead to larger harvests, but the faster-growing cohorts are harvested first, so that later harvests are depressed when serial correlation continues for longer durations. In this particular case, these consequences are minor, as the model was enhanced and refined (Vanclay 1994a), and timber harvesting ceased (Vanclay 1996a), but this example highlights the importance of comprehensive sensitivity testing of all model components. The lesson from this study is not the importance of serial correlation (which may be less in models that use basal area in larger trees to predict increment, Ledermann and Eckmüllner, 2004), but of the possible consequences of untested assumptions.

This re-appraisal of an established simulator should remind readers of the importance of sensitivity testing of assumptions, both explicit and implicit. Clearly, the constructs offered in software, the operating system and the hardware available may all influence the representation, implementation and performance of a simulator. It remains instructive for all modellers to constantly question whether a representation is faithful to their imagination or slave to the technology available, and to be aware that seemingly innocuous assumptios may have significant consequences for predictions.

*5. Discussion*

Vanclay's (1989a) model was a proof-of-concept that signalled a change from long-established stand-table projection approaches (Higgins, 1977) to dynamic simulation and yield scheduling (Vanclay, 1990, 1994a). Most of the embedded functional relationships were subsequently enhanced: site quality assessment (Vanclay, 1989b), diameter increment (Vanclay, 1991a), mortality (Vanclay, 1991b), regeneration (Vanclay, 1992), merchantability (Vanclay, 1991c) and harvesting (Vanclay, 1989c) functions were all subsequently revised to include more species groups and more variables. However, the prototype simulator presented here in Simile offers pedagogic advantages as it embodies the design and structure of the operational version, without the additional complexity of many equations with multiple parameters (i.e., hundreds of species and dozens of equations, each with several parameters). Despite its comparative simplicity, the simulator involves assumptions that appeared innocuous in the Fortran implementation, but are now revealed in Simile to have substantial consequences for resulting predictions.

This practical example illustrates that the quality of a model cannot be judged independently of its implementation as a simulator, and that apparently minor assumptions made in implementing the simulator may have substantial influence on predictions. A motor car offers a familiar and pedagogic analogy: a perfect engine cannot perform well in an unroadworthy car, and the utility of the vehicle relies on its overall performance rather than its theoretical specifications. So it is with simulation models: the utility of a model depends in part on assumptions made during model implementation as a simulator, and it behoves modellers to reveal, document and test such assumptions. This requires some thoughtful reflection by the modelling team, as familiarity with a modelling environment may lead to the intuitive use of convenient constructs (e.g., arrays in Fortran; submodels in Simile) without a full appreciation of the consequences for predictions.

Other related experiences are also insightful: several colleagues have related anecdotes in which thoughtful bounding to avoid computation problems in a model (such as Y=max($\boldsymbol{\beta}\mathbf{X}$, $c$)) has resulted in the use of the upper bound ($c$) in most simulations, rather than the intended function $\boldsymbol{\beta}\mathbf{X}$. Sadly, many such discoveries are discovered only accidentally and belatedly, and they are rarely reported in the literature.

This Simile representation makes the model accessible for teaching purposes, and encourages exploration into the consequences of implementation decisions such as record splitting that have received little consideration. Similar experiences have been reported during re-engineering of the landscape model LANDIS (Scheller *et al.*, 2010). Others have shared insights gained from converting a simulator from one computer language to another (e.g., Cumming and Burton, 1993), or to a hybrid implementation combining both code-based and icon-based software (e.g., Smith *et al.*, 2005; Lättilä *et al.*, 2010), but such conversion appears to offer fewer insights than re-engineering or recasting a simulator in visual environments.

*6. Conclusion*

Familiarity with a modelling platform, whether Fortran, Simile or otherwise, can be seductive, and can lure modellers into introducing untested assumptions for convenience of implementation. As demonstrated in this paper, such assumptions may have consequences greater than suspected. Modellers should remain conscious of all assumptions, should consider alternative implementations that make assumptions evident, and should conduct sensitivity tests to inform decisions.

*References*

Alder, D., Silva, J.N.M., 2000. An empirical cohort model for management of *Terra Firme* forests in the Brazilian Amazon. *Forest Ecology and Management* 130, 141-157.
Anonymous, 1981. Rainforest Volume Equations. Research Report No 3, Department of Forestry, Queensland (ISSN 0311-0893), pp. 85.
Anonymous, 1987. Rainforest increment studies. Research Report No 5, Department of Forestry, Queensland (ISSN 0311-0893), pp. 59-60.
Blackwell, G.L., Potter, M.A., Minot, E.O., 2001. Rodent and predator population dynamics in an eruptive system. *Ecological Modelling* 142(3), 227-245.
Bossel, H., 1991. Modelling forest dynamics: Moving from description to explanation. *Forest Ecology and Management* 42, 129–142.
Botkin, D.B., 1993. *Forest Dynamics: an ecological model*. Oxford University Press, Oxford. 309 pp.
Brown, W.J., Malveau, R.C., McCormick, H.W., Mowbray, T.J., 1998. *AntiPatterns: Refactoring Software, Architectures, and Projects in Crisis*. Wiley.
Buckman, R.E., 1962. Growth and yield of red pine in Minnesota. USDA For. Serv., Tech. Bull. 1272.
Clark, D.A., Clark, D.B., 1999. Assessing the growth of tropical rain forest trees: Issues for forest modeling and management. *Ecological Applications* 9, 981–997.
Clutter, J.L., 1963. Compatible growth and yield models for loblolly pine. *Forest Science* 9(3), 354-371.
Cole, D. M., Stage, A. R. 1972. Estimating future diameters of lodgepole pine. USDA For. Serv. Res. Pap. INT-131, 20 p. Intermt. For. and Range Exp. Stn., Ogden Utah.
Connell, J.H., Green, P.T., 2000. Seedling dynamics over thirty-two years in a tropical rain forest tree. *Ecology* 81, 568–584.
Crookston, N.L., Dixon, G.E., 2005. The forest vegetation simulator: A review of its structure, content, and applications. *Computers and Electronics in Agriculture* 49, 60–80.
Cumming, S.G., Burton, P.J., 1993. A programmable shell and graphics system for forest stand simulation. *Environmental Software* 8, 219–230.

Davey, C., Ougham, H.J., Millar, A., Thomas, H., Tindal, C., Muetzelfeldt, R., 2009. PlaSMo: Making existing plant and crop mathematical models available to plant systems biologists. *Comparative Biochemistry and Physiology - Part A: Molecular & Integrative Physiology* 153, S225–S226.

Doerr, H. M., 1996. Stella ten years later: A review of the literature. *International Journal of Computers for Mathematical Learning* 1(2), 201-224.

Dufour-Kowalski, S., Courbaud, B., Dreyfus, P., Meredieu, C., de Coligny, F., 2012. Capsis: an open software framework and community for forest growth modelling. *Annals of Forest Science* 69, 221–233.

Duursma, R.A., Robinson, A.P., 2003. Bias in the mean tree model as a consequence of Jensen's inequality. *Forest Ecology and Management* 186, 373–380.

Elkin, C., Reineking, B., Bigler, C., Bugmann, H., 2012. Do small-grain processes matter for landscape scale questions? Sensitivity of a forest landscape model to the formulation of tree growth rate. *Landscape Ecology* 27(5), 697-711.

Eskrootchi, R., & Oskrochi, G. R., 2010. A Study of the Efficacy of Project-based Learning Integrated with Computer-based Simulation - STELLA. *Educational Technology & Society* 13 (1), 236–245.

Ford, A., 1999. *Modeling the environment: an introduction to system dynamics modeling of environmental systems.* Island Press.

Garcia, O., 2013. Forest Stands as Dynamical Systems: An Introduction. *Modern Applied Science* 7(5), 32-38.

Garcia, O., Ruiz, F., 2003. A growth model for eucalypt in Galicia, Spain. *Forest Ecology and Management* 173, 49–62.

Haggith, M., Prabhu, R., Pierce Colfer, C.J., Ritchie, B., Thomson, A., Mudavanhu, H., 2003. Infectious Ideas: Modelling the Diffusion of Ideas across Social Networks. *Small-scale Forestry* 2, 225-239.

Hamilton, D.A., Jr., 1994. Uses and abuses of multipliers in the Stand Prognosis Model. Gen. Tech. Rep. INT-GTR-310. Ogden, UT: U.S. Department of Agriculture, Forest Service, Intermountain Research Station. 9 p.

Hedayat, A.S., Su, G., 2012. Robustness of the simultaneous estimators of location and scale from approximating a histogram by a normal density curve. *The American Statistician* 66(1), 25-33.

Higgins, M.D., 1977. A sustained yield study of north Queensland rainforests. Queensland Department of Forestry, 182 pp.

Jacoby, J. E., Harrison, S., 1962. Multi-variable experimentation and simulation models. *Naval Research Logistics Quarterly* 9(2), 121-136.

Jensen, J.L., 1906. Sur les fonctions convexes et les inégualités entre les valeurs moyennes. *Acta Mathematica* 30, 175–193.

Kalgraf, K., 1979. The dynamics of a simple stand. In L. Lönnstedt and J. Randers (eds) Wood resource dynamics in the Scandinavian forestry sector. *Studia Forestalia Suecica* 152, 55-66.

Kariuki, M., Kooyman, R.M., Smith, R.G.B., Wardell-Johnson, G., Vanclay, J.K., 2006. Regeneration changes in tree species abundance, diversity and structure in logged and unlogged subtropical rainforest over a thirty six year period. *Forest Ecology and Management* 236, 162-176.

Köhler, P., Huth, A., 1998. The effects of tree species grouping in tropical rain forest modelling: Simulations with the individual based model Formind. *Ecological Modelling* 109, 301-321.

Lättilä, L., Hilletofth, P., Lin, B., 2010. Hybrid simulation models–when, why, how? *Expert Systems with Applications* 37(12), 7969-7975.

Leary, R.A., 1985. *Interaction Theory in Forest Ecology and Management*. Nijnhoff, Dordrecht, 219 pp.

Ledermann, T., Eckmüllner, O., 2004. A method to attain uniform resolution of the competition variable Basal-Area-in-Larger Trees (BAL) during forestgrowth projections of small plots. *Ecological Modelling* 171, 195-206.

Muetzelfeldt, R., 2004. Position paper on declarative modelling in ecological and environmental research. European Commission Directorate-General for Research, Sustainable development, global change and ecosystems. ISBN 92-894-5212-9.

Muetzelfeldt, R., 2010. A generic approach for representing complex structures in biological models. *Nature Precedings* DOI 10.1038/npre.2010.5188.1

Muetzelfeldt, R., Robertson, D., Bundy, A., Uschold, M., 1989. The use of Prolog for improving the rigour and accessibility of ecological modelling. *Ecological Modelling* 46, 9-34.

Muetzelfeldt, R., Taylor, J., 1997. The suitability of AME (the Agroforestry Modelling Environment) for agroforestry modelling. *Agroforestry Forum* 8, 2, 7-9.

Muetzelfeldt, R., Massheder, J., 2003. The Simile visual modelling environment. *European Journal of Agronomy* 18, 345-358.

Nebel, G., Dragsted, J., Vanclay, J.K., 2001. Structure and floristic composition of flood plain forests in the Peruvian Amazon: II. The understorey of restinga forests. *Forest Ecology and Management* 150, 59-77.

Okuda, T., Kachi, N., Yap, S.K., Manokaran, N.,1997. Tree distribution pattern and fate of juveniles in a lowland tropical rain forest – implications for regeneration and maintenance of species diversity. *Plant Ecology* 131, 155–171.

Ong, P.C., Kleine, M., 1996. DIPSIM: Dipterocarp forest growth simulation model – a tool for forest-level management planning. In A. Schulte (ed. ) *Dipterocarp Forest Ecosystems: Towards sustainable management*, World Scientific, pp. 228-246.

Picard, N., Franc, A., 2003. Are ecological groups of species optimal for forest dynamics modelling? *Ecological Modelling* 163, 175-186.

Porte, A., Bartelink, H.H., 2002. Modelling mixed forest growth: a review of models for forest management. *Ecological Modelling* 150, 141–188.

Prabhu, R., Haggith, M., Mudavanhu, H., Muetzelfeldt, R., Standa-Gunda, W., Vanclay, J.K., 2003. ZimFlores: A model to advise co-management of the Mafungautsi Forest in Zimbabwe. *Small-scale Forestry* 2, 185-210.

Preston, R.A., Vanclay, J.K., 1988. Calculation of timber yields from north Queensland rainforests. Queensland Department of Forestry, Technical Paper No 47. 19 p.

Pretzsch, H., Biber, P., Dursky, J., von Gadow, K., Hasenauer, H., Kändler, G., Kenk, G., Kublin, E., Nagel, J., Pukkala, T., Skovsgaard, J.P., Sodtke, R., Sterba, H., 2002. Recommendations for standardized documentation and further development of forest growth simulators. *Forstw. Cbl.* 121, 138-151.

Reineke, L.H., 1927. A modification of Bruce's method of preparing timber yield tables. *J. Agric. Res.* 35, 843–856.

Robinson, A.P., Monserud, R.A., 2003. Criteria for comparing the adaptability of forest growth models. *Forest Ecology and Management* 172, 53-67.

Scheller, R.M., Sturtevant, B.R., Gustafson, E.J., Ward, B.C., Mladenoff, D.J., 2010. Increasing the reliability of ecological models using modern software engineering techniques. *Frontiers in Ecology and the Environment* 8, 253–260.

Simon, G., 1976. Computer simulation swindles, with applications to estimates of location and dispersion. *Appl. Statist.* 25, 266–274.

Simulistics, 2013. Simile at a glance. Simulistics Ltd. http://www.simulistics.com/overview.htm [16 May 2013].

Smith, F.P., Holzworth, D.P., & Robertson, M.J., 2005. Linking icon-based models to code-based models: a case study with the agricultural production systems simulator. *Agricultural Systems* 83(2), 135-151.

Stage, A.R., 1973. Prognosis model for stand development. Res. Pap. INT-137. USDA Forest Service, Intermountain Forest and Range Experiment Station. 32 p.

Valle, D., 2011. Incorrect representation of uncertainty in the modeling of growth leads to biased estimates of future biomass. *Ecological Applications* 21(4), 1031-1036.

Vanclay, J.K., 1983. Techniques for modelling timber yield from indigenous forests with special reference to Queensland. M.Sc. Thesis, University of Oxford, U.K., 194 p.

Vanclay, J.K., 1989a. A growth model for north Queensland rainforests. *Forest Ecology and Management* 27, 245-271.

Vanclay, J.K., 1989b. Site productivity assessment in rainforests: an objective approach using indicator species. In: Wan Razali Mohd, H.T. Chan and S. Appanah (eds) Proceedings of the Seminar on Growth and Yield in Tropical Mixed/Moist Forests, 20-24 June 1988, Kuala Lumpur. Forest Research Institute Malaysia, p. 225-241.

Vanclay, J.K., 1989c. Modelling selection harvesting in tropical rain forests. *Journal of Tropical Forest Science* 1, 280-294.

Vanclay, J.K., 1990. Design and implementation of a state-of-the-art inventory and forecasting system for indigenous forests. In: H.G. Lund and G. Preto (eds) Global Natural Resource Monitoring and Assessment: Preparing for the 21st century, Venice, Italy, 24-30 September 1989. Pp. 1072-1078.

Vanclay, J.K., 1991a. Compatible deterministic and stochastic predictions by probabilistic modelling of individual trees. *Forest Science* 37, 1656-1663.

Vanclay, J.K., 1991b. Mortality functions for north Queensland rainforests. *Journal of Tropical Forest Science* 4, 15-36.

Vanclay, J.K., 1991c. Modelling changes in the merchantability of individual trees in tropical rainforest. *Commonwealth Forestry Review* 70, 105-111.

Vanclay, J.K., 1991d. Data requirements for developing growth models for tropical moist forests. *Commonwealth Forestry Review* 70, 248-271.

Vanclay, J.K., 1992. Modelling regeneration and recruitment in a tropical rainforest. *Canadian Journal of Forest Research* 22, 1235-1248.

Vanclay, J.K., 1994a. Sustainable timber harvesting: Simulation studies in the tropical rainforests of north Queensland. *Forest Ecology and Management* 69, 299-320.

Vanclay, J.K., 1994b. *Modelling Forest Growth and Yield: Applications to Mixed Tropical Forests*. CAB International, Wallingford, U.K.

Vanclay, J.K., 1996a. Lessons from the Queensland rainforests: Steps towards sustainability. *Journal of Sustainable Forestry* 3(2/3):1-27.

Vanclay, J.K., 1996b. Assessing the sustainability of timber harvests from natural forests: Limitations of indices based on successive harvests. *Journal of Sustainable Forestry* 3(4):47-58.

Vanclay, J.K., 2003. Realizing opportunities in forest growth modelling. *Canadian Journal of Forest Research* 33, 536-541.

Vanclay, J.K., 2006a. Spatially-explicit competition indices and the analysis of mixed-species plantings with the Simile modelling environment. *Forest Ecology and Management*, 233, 295-302.

Vanclay, J.K., 2006b. Can the lessons from the Community Rainforest Reforestation Program in eastern Australia be learned? *International Forestry Review* 8(2):256-264.

Villa, F., Athanasiadis, I.N., Rizzoli, A.E., 2009. Modelling with knowledge: A review of emerging semantic approaches to environmental modelling. *Environmental Modelling &Software* 24, 577-587.

Weiskittel, A.R., D.W. Hann, J.A. Kershaw and J.K. Vanclay, 2011. *Forest Growth and Yield Modeling*. Wiley.

Wirth, N., 1985. *Algorithms + Data Structures=Programs* (Vol. 76). Prentice Hall.